\documentclass[12pt]{article}

\usepackage{comment}
\usepackage{tikz}
\usepackage{caption}
\usepackage{bigints}

\usepackage{mathtools}

\usepackage[mathscr]{euscript}

\usepackage{slashed}

\usepackage{amsmath}

\date{}

\title{\textbf{General Solution and Canonical Quantization of the
Conic Path Constrained Second-Class System
}}
\author{ \textbf{R. L. Caires, S. L. Oliveira and R. Thibes${}^1$}
\\\\
\textit{\small{Departamento de Ci\^encias Exatas e Naturais}},\\
\textit{\small{Universidade Estadual do Sudoeste da Bahia}},\\
\textit{\small{Rodovia BR 415, km 03, s/n - Itapetinga - BA}},
\textit{\small{45700-000 Brazil}}\\
\textit{\small{${}^1$thibes@uesb.edu.br}}
 }

\usepackage{graphicx}
\usepackage{amssymb,amsfonts,amssymb,amsthm}
\usepackage{amsmath}

\begin{document}

\maketitle

\abstract{We consider the problem of constrained motion along a conic path under a given external potential function.  The model is described as a second-class system capturing the behavior of a certain class of specific quantum field theories.  By exhibiting a suitable integration factor,
we obtain the general solution for the associated non-linear differential equations.  We perform the canonical quantization in a consistent way in terms of the corresponding Dirac brackets.
We apply the Dirac-Bergmann algorithm to unravel and classify the whole internal constraints structure inherent to its dynamical Hamiltonian description, obtain the proper extended Hamiltonian function, determine the Lagrange multiplier and  compute all relevant Poisson brackets among the constraints, Hamiltonian and Lagrange multiplier. The complete Dirac brackets algebra in phase space as well as its physical realization in terms of differential operators is explicitly obtained.}

{\bf keywords}: singular systems; Dirac-Bergmann algorithm; Dirac brackets; constrained systems; conic particle; quantization methods

\section{Introduction}
Quantum field theory is allegedly the most fundamental framework to accurately describe the deepest subtleties of nature phenomena concerning the known physical world.  This framework includes both the very successful standard model of elementary particles and its possible extensions, as well as the current tentative proposals for understanding quantum gravity, passing through supersymmetry, string theory and loop quantum gravity. 
It is possible to obtain enlightening insights about quantum field theory by exploring specific features of crafty properly designed quantum mechanical models.  As a short sample, we mention the references
\cite{Plyushchay:1990kg, Loeffelholz:1996cv,  Shimizu:2005mq, Shukla:2015wka, Barbosa:2018dmb, Barbosa:2018bng, deOliveira:2019eva, Rizzuti:2021xio, Pandey:2020amp}.  In \cite{Plyushchay:1990kg} we see an alternative mechanical model for spinning particles, while in \cite{Loeffelholz:1996cv} a quantum mechanical gauge model is constructed from a particle on a circle mimicking the massive Schwinger model in order to discuss the Strong CP Problem.  A free particle in $S^D$ is studied in \cite{Shimizu:2005mq} via discretized BRST\footnote{Becchi-Rouet-Stora-Tyutin \cite{Becchi:1974xu, Becchi:1974md, Tyutin:1975qk}.} transformations leading to the evaluation of the corresponding quantum mechanical potential.  A supervariable approach to a supersymmetric quantum mechanical model can be seen in \cite{Shukla:2015wka}.  A free particle moving along a conic curve has been discussed in references \cite{Barbosa:2018dmb, Barbosa:2018bng} whilst a more general Hamiltonian mechanical model with gauge invariance, treated via the Dirac-Bergmann constraints algorithm, can be seen in \cite{deOliveira:2019eva}.  Connections between field theory and classical mechanics through the gauge principle as a local symmetry have been investigated in \cite{Rizzuti:2021xio}.  The BRST quantization of a particle in a Riemannian manifold can be seen in \cite{Pandey:2020amp}. In those and many other similar works, one obtains important clues from mechanical models which can be used to approach more challenging issues in quantum field theory.  Additionally, quantum mechanical models can also be used for didatical purposes in quantum field theory, as can be seen for instance in \cite{Nemeschansky:1987xb} where a particle on a circle is used to explain BRST symmetry and the role of ghost fields in QED\footnote{Quantum Electrodynamics} and QCD\footnote{Quantum Chromodynamics}.

Along this line, in the current work, we consider a particle moving in an arbitrary conic path under the action of a given external potential function as  a Dirac-Bergmann constrained dynamical system, providing a thoroughly detailed analysis concerning both its Lagrangian and Hamiltonian descriptions and perform its canonical quantization. The problem is characterized as a singular Dirac-Bergmann second-class system governed by coupled non-linear differential equations whose general solution is obtained here in full detail.  Some aspects of a simplified version of this model have been considered in \cite{Barbosa:2018dmb, Barbosa:2018bng}, which actually generalize a few ideas already seen in \cite{Nemeschansky:1987xb}.  Aiming at a direct analogy with quantum chromodynamics, reference \cite{Nemeschansky:1987xb} elaborates on different possible descriptions for the motion of a particle along a circular path as a constrained system.  The Lagrange multiplier used to implement the constraint acquires dynamics and plays a similar role as the time component of the gauge field $A^a_0$ \cite{Nemeschansky:1987xb}.  The Faddeev-Popov ghosts of QCD are shown to have a similar counterpart in the mechanical model which also enjoys a natural BRST symmetry.  The main raw ideas present in \cite{Nemeschansky:1987xb}, some of them introduced in an {\it ad hoc} manner, were further justified, explained, generalized and organized in a systematic way in references \cite{Barbosa:2018dmb, Barbosa:2018bng}.  Starting from two consistent Lagrangian and Hamiltonian mechanical models, the BRST quantization of a conic constrained particle is fully implemented in \cite{Barbosa:2018dmb} whereas a symplectic approach via FJBW\footnote{Faddeev-Jackiw-BarcelosNeto-Wotzasek \cite{Faddeev:1988qp, Barcelos-Neto:1991rxi}.} can be seen in \cite{Barbosa:2018bng}.  Despite being constructed initially for comparative purposes, the model introduced in references \cite{ Barbosa:2018dmb, Barbosa:2018bng, Nemeschansky:1987xb} grew up and acquired interest in itself being subsequently used as a starting point for further discussions regarding (anti-)BRST symmetry \cite{Gupta:2009dy}, superfield formalism \cite{Shukla:2014spa} and BFFT\footnote{Batalin-Fradkin-Fradkina-Tyutin \cite{Batalin:1986aq, Batalin:1989dm, Batalin:1991jm}.} Abelian conversion \cite{Pandey:2021myh}, giving rise to interesting applications in field theory such as the Lorentz breaking bumblebee model \cite{Pandey:2021myh} and the non-linear sigma model \cite{Thibes:2020yux}.

Concerning the two previous mentioned works \cite{Barbosa:2018dmb} and \cite{Barbosa:2018bng}, the present letter reports new important facts regarding the inclusion of a general open potential function, the complementation of missing key calculations such as the Lagrange multiplier function and the whole Poisson brackets (PB) algebra among all constraints, Lagrange multiplier and Hamiltonian function and, most importantly, the obtainment of the general classical solution for the equations of motion and its complete canonical quantization.  In fact, after introducing a general potential function responsible for a richer and broader dynamics, the system evolution is no longer restrict to geodesics.  With suitable algebraic steps, by means of an elimination of variables procedure, we obtain a non-linear second order differential equation which we show can be solved in terms of a suitable integration factor.
For the reader's convenience, this article is organized as follows.  In section {\bf 2} below, we characterize the conic path model  as a second-class constrained system and apply the Dirac-Bergmann algorithm to compute the complete constraints set.  The initial conic constraint equation is imposed by means of an extra variable in configuration space acting as an undetermined Lagrange multiplier.  Aiming for more generality, we allow for an open external potential function providing a nontrivial dynamics for the system.  We compute the Lagrange multiplier as well as all Poisson brackets among it and the constraints in full form, showing the stability of the constraints surface within phase space.  In section {\bf 3}, we discuss the equations of motion in detail.  We obtain the main non-linear second order differential equation for the second-class system and its solution through the mentioned integration factor.  To illustrate better the ideas, we work out a particular case as an example for an elliptical trajectory.  The canonical quantization of the system is the subject of section {\bf 4}, in which we compute all fundamental Dirac brackets (DB) of the theory and present a differential operators realization of the Dirac algebra at quantum level.  We conclude in section {\bf 5} with some final remarks.  Two appendixes {\bf A} and {\bf B} are reserved to technical calculations supporting some results of sections {\bf 2} and {\bf 4} respectively.

\section{The Singular Second-Class System}
Consider a mass $m$ particle
constrained to move along a non-degenerated conic path of the form
\begin{equation}\label{gcon}
Ax^2+Bxy+Cy^2+Dx+Ey+F=0
\,,
\end{equation}
where the capital letters $A,\dots,F$ stand for real numbers denoting the specific conic parameters while $(x,y)$ are usual Cartesian coordinates.
Associated to (\ref{gcon}), we define the conic discriminant
\begin{equation}
\Gamma\equiv B^2-4AC\,,
\end{equation}
whose signal
characterizes an elliptical ($\Gamma < 0$), parabolic ($\Gamma = 0$) or hyperbolic ($\Gamma >0$) path.

The restriction to the conic path can be implemented by means of a third coordinate $z$ used as a tool to generate relation (\ref{gcon}) as one of the equations of motion resulting from a variational principle.  In this way, a Lagrangian function for this system can be written as
\begin{equation}\label{L}
L = \dfrac{m}{2}(\dot{x}^2+\dot{y}^2)-V(x,y)-z(Ax^2+Bxy+Cy^2+Dx+Ey+F)
\,,
\end{equation} 
with $V(x,y)$ denoting a given potential function responsible for the dynamics of the particle motion along the conic (\ref{gcon}).  This is a dynamical system described by the three coordinates $x$, $y$ and $z$ as functions of a real parameter $t$ corresponding to the physical time.

Actually, the Lagrangian (\ref{L}) represents a {\it constrained} dynamical system in the sense of Dirac and Bergmann \cite{Dirac:1950pj, Anderson:1951ta, Dirac}.  This can be seen by defining the usual canonical momenta as
\begin{equation}
p_x=\frac{\partial L}{\partial\dot{x}}=m\dot{x}\,,\,\,\,\,\,
p_y=\frac{\partial L}{\partial\dot{y}}=m\dot{y}\,,\,\,\,\,\,
p_z=\frac{\partial L}{\partial\dot{z}}=0
\end{equation}
and observing that $\chi_1\equiv p_z$ corresponds to a primary constraint in phase space.  As a consequence, the Legendre transformation from configuration to phase space is not invertible.
The canonical Hamiltonian corresponding to (\ref{L}) can be worked out as
\begin{equation}
H_c=\frac{1}{2m}(p_x^2+p_y^2)+V(x,y)+z(Ax^2+Bxy+Cy^2+Dx+Ey+F)
\,.
\end{equation}
To unravel all phase space internal constraints of the model, following the Dirac-Bergmann prescription
\cite{Dirac:1950pj, Anderson:1951ta, Dirac, Sundermeyer:1982gv, Gitman:1990qh, Henneaux:1992ig}, we introduce a Lagrange multiplier function $\lambda$,
define the primary Hamiltonian as
\begin{equation}\label{HP}
H_P = H_c + \lambda p_z
\,,
\end{equation}
and impose the time conservation of the primary constraint $\chi_1$ and its subsequent secondary ones. 
Explicitly, the time stability condition for $\chi_1$ can be obtained by demanding
\begin{equation}
\{\chi_1,H_P\}=-Ax^2-Bxy-Cy^2-Dx-Ey-F=0
\end{equation}
which leads to the secondary constraint\footnote{The sign choice for $\chi_2$ is of course just a matter of convention.}
\begin{equation}
\chi_2\equiv Ax^2+Bxy+Cy^2+Dx+Ey+F\,.
\end{equation}
Proceeding this way further with the Dirac-Bergmann algorithm, two more constraints can be found, namely
\begin{equation}
\{\chi_2,H_P\}=\frac{1}{m}
\big[ \left(2Ap_x+Bp_y\right)x+\left(Bp_x+2Cp_y\right)y+Dp_x+Ep_y \big]\equiv \chi_3 \,,
\end{equation}
and
\begin{equation}
\{\chi_3,H_P\}=\left[Ap_x^2+Bp_xp_y+Cp_y^2\right]-\frac{z}{m}M(x,y)-\frac{1}{m}N(x,y)\equiv\chi_4
\,,
\end{equation}
with $M(x,y)$ and $N(x,y)$  defined for convenience as
\begin{equation}\label{M}
M(x,y)
\equiv\left[(2Ax+By+D)^2+(Bx+2Cy+E)^2\right]
\,,
\end{equation}
and
\begin{equation}
N(x,y)\equiv
\left[ (2Ax+By+D)V_x(x,y) + (Bx+2Cy+E)V_y(x,y) \right]
\,,
\end{equation}
where
\begin{equation}
V_x\equiv\frac{\partial V}{\partial x}\mbox{ ~~and~~ } V_y\equiv\frac{\partial V}{\partial y}\,.
\end{equation}
The conservation of $\chi_4$ does not lead to new constraints, but rather determines the Lagrange multiplier introduced in equation (\ref{HP}) as a function of the phase space variables given by
\begin{eqnarray}\label{lambda}
\lambda&=&-{(mM)}^{-1}\bigg\lbrace
(2Ax+By+D)\Big[
p_x(V_{xx}+8Az)+p_y(V_{xy}+4Bz)\Big]
\nonumber\\&&
+(Bx+2Cy+E)\Big[
p_x(V_{xy}+4Bz)+p_y(V_{yy}+8Cz)\Big]
\nonumber\\&&
+3V_y\left(Bp_x+2Cp_y\right)
+3V_x\left(2Ap_x+Bp_y\right)
\bigg\rbrace
\,.
\end{eqnarray}

Summarizing, in phase space, the system (\ref{L}) possesses a chain of four constraints $\chi_{r}$, with $r=1,\dots,4$, given by
\begin{equation}\label{constraints}
\begin{aligned}
\chi_1&=p_z\,,
\,\,\,\,\,\,
\chi_2=Ax^2+Bxy+Cy^2+Dx+Ey+F\,,
\\
\chi_3&=\frac{1}{m}
\big[ \left(2Ap_x+Bp_y\right)x+\left(Bp_x+2Cp_y\right)y+Dp_x+Ep_y \big]\,,
\\
\chi_4&=\frac{2}{m^2}
\left[Ap_x^2+Bp_xp_y+Cp_y^2\right]-\frac{z}{m}M(x,y)-\frac{1}{m}N(x,y)
\,.
\end{aligned}
\end{equation}
The substitution of (\ref{lambda}) back into (\ref{HP}) produces then the final Hamiltonian
\begin{equation}\label{H}
H(x,y,z,p_x,p_y,p_z)=H_P(x,y,z,p_x,p_y,p_z,\lambda(x,y,z,p_x,p_y))
\,,
\end{equation}
under which the constraints (\ref{constraints}) remain stationary within the phase space constraints hypersurface.  
For completeness and to pave the way for calculating the Dirac brackets among the dynamical quantities, including all coordinates and their conjugate momenta, we compute next the PBs between all constraints and the primary Hamiltonian and write down the results in full form.
In fact, evaluating the PBs between $H$ and the constraints, we obtain the structure
\begin{equation}\label{Halgebra}
\begin{aligned}
\{\chi_1,H\}&=-\chi_2+\{\chi_1,\lambda\}\chi_1\,,\\
\{\chi_2,H\}&=\chi_3+\{\chi_2,\lambda\}\chi_1\,,\\
\{\chi_3,H\}&=\chi_4+\{\chi_3,\lambda\}\chi_1\,,\\
\{\chi_4,H\}&=\{\chi_4,\lambda\}\chi_1\,,
\end{aligned}
\end{equation}
with
\begin{eqnarray}
\{\chi_1,\lambda\}&=&4{(mM)}^{-1}
\bigg\lbrace
\left(2Ap_x+Bp_y\right)\left(2Ax+By+D\right)
\nonumber\\&&
+
\left(Bp_x+2Cp_y\right)\left(Bx+2Cy+E\right)
\bigg\rbrace
\,,
\end{eqnarray}
\begin{eqnarray}
\{\chi_2,\lambda\}&=&
-{(mM)}^{-1}
\bigg\lbrace
(2Ax+By+D)^2(V_{xx}+8Az)+(Bx+2Cy+E)^2(V_{yy}+8Cz)
\nonumber\\&&
+3(2Ax+By+D)(2AV_x+BV_y)+3(Bx+2Cy+E)(2CV_y+BV_x)
\nonumber\\&&
+2(V_{xy}+4Bz)(2Ax+By+D)(Bx+2Cy+E)
\bigg\rbrace
\,,
\end{eqnarray}
\begin{equation}\label{chi3lambda}
\{\chi_3,\lambda\}=
-\frac{1}{m^2M^2}\left[
p_x \Omega^{x}(x,y,z) + p_y \Omega^{y}(x,y,z)
\right]
\,,
\end{equation}
and
\begin{eqnarray}\label{chi4lambda}
\{\chi_4,\lambda\}&=&
p_x^2F_{(2)}^{xx}+p_xp_yF_{(2)}^{xy}+p_y^2F_{(2)}^{yy}
+F_{(0)}
\,.
\end{eqnarray}
The functions $\Omega$ and $F$ appearing in (\ref{chi3lambda}) and (\ref{chi4lambda}) do not depend on the momenta and are calculated in more detail in Appendix {\bf A}.  The set of equations (\ref{Halgebra}) assures the constraints stability and consistency of the system.

On the other hand, the constraints (\ref{constraints}) do not close an algebra among themselves, but rather produce the following non-null  PB relations:
\begin{equation}\label{CPB14}
\{\chi_1,\chi_4\}=\{\chi_2,\chi_3\}=m^{-1}M(x,y)
\,,
\end{equation}
\begin{eqnarray}\label{CPB24}
\{\chi_2,\chi_4\}&=&\dfrac{2}{m^2}[(2Ax+By+D)(2Ap_x+Bp_y)
\nonumber\\&&
+(Bx+2Cy+E)(2Cp_y+Bp_x)]=\frac{M(x,y)\{\chi_1,\lambda\}}{2m}
\,,
\end{eqnarray}
and
\begin{equation}
\begin{aligned}\label{CPB34}
\{\chi_3,\chi_4\}=&
\dfrac{2}{m^3}\Big[ (2Ap_x+Bp_y)^2+(2Cp_y+Bp_x)^2\Big]
\\&
+\dfrac{1}{m^2}\Big[(V_{xx}+4zA)(2Ax+By+D)^2 + (V_{yy}+4zC)(Bx+2Cy+E)^2
\\&
 +(2V_{xy}+4zB)(2Ax+By+D)(Bx+2Cy+E)
\\&
+ (2Ax+By+D)(2AV_x+BV_y)+(Bx+2Cy+E)(BV_x+2CV_y)\Big]
\,.
\end{aligned}
\end{equation}

As consequence, the PB algebra given by the set of equations (\ref{Halgebra}) and (\ref{CPB14}) to (\ref{CPB34}) above characterize a second-class system, according to the well-known Dirac-Bergmann constraints classification \cite{Dirac:1950pj, Anderson:1951ta, Dirac, Sundermeyer:1982gv, Gitman:1990qh, Henneaux:1992ig}.  This can be seen by investigating the determinant of the constraint matrix constructed from the previous PB relations.  A direct calculation shows that
\begin{equation}\label{det}
\det \{\chi_r,\chi_s\} = m^{-4}M^4 
\,.
\end{equation}
Since (\ref{det}) is not zero, the system is indeed second-class and, particularly, does not enjoy gauge invariance.\footnote{Gauge invariance is related to first class constraints \cite{Henneaux:1992ig, Thibes:2020jfp}.  An alternative description of the present system with gauge symmetry can be done along the lines of reference \cite{deOliveira:2019eva}.} 

\section{General Solution}
The Hamiltonian formulation will prove essential for the canonical quantization of the model.  Concerning its classical development however, in order to obtain a direct and clear general solution in a sound incontestable way, in this section we come back to its Lagrangian description.
The two Euler-Lagrange equations associated to (\ref{L}) corresponding to the variables $x$ and $y$ can be respectively written as
\begin{equation}
\begin{aligned}
m\ddot{x}+z\left(2Ax+By+D\right)&=-V_x\,,\\
m\ddot{y}+z\left(Bx+2Cy+E\right)&=-V_y\,
\end{aligned}
\end{equation}
or, equivalently,
\begin{equation}\label{z}
z=-\frac{V_x+m\ddot{x}}{2Ax+By+D}=-\frac{V_y+m\ddot{y}}{Bx+2Cy+E}\,.
\end{equation}
Upon eliminating the auxiliary variable $z$, we obtain the mixed differential equation
\begin{equation}
\left(Bx+2Cy+E\left)\right(m\ddot{x}+V_x\right)
=
\left(2Ax+By+D\right)\left(m\ddot{y}+V_y\right)
\,,
\end{equation}
relating the two unknown functions $x(t)$ and $y(t)$.  Furthermore, using (\ref{gcon}), we may eliminate one of these last two, let us say $y(t)$, in terms of the other and
write an autonomous non-linear second-order ordinary differential equation for $x(t)$ in the form
\begin{equation}\label{ODE}
P_{\pm}(x)\ddot{x}+Q_{\pm}(x)\dot{x}^2+R_{\pm}(x)=0\,,
\end{equation}
with $P_{\pm}(x)$, $Q_{\pm}(x)$ and $R_{\pm}(x)$ given by
\begin{equation}\label{P}
P_{\pm}(x)=\pm J^{1/2}\left[
1+\frac{(B\mp J^{-1/2}K)^2}{4C^2}
\right]
\,,
\end{equation}
\begin{equation}\label{Q}
Q_{\pm}(x)=\pm J^{-1/2}\,\frac{\left(K \mp J^{1/2}B \right)\left(\Gamma-K^2J^{-1}\right)}{4C^2}
\,,
\end{equation}
\begin{equation}\label{R}
R_{\pm}(x)=\frac{1}{m}\left[
\frac{\left(K \mp J^{1/2}B \right)}{2C}V_y(x) \pm J^{1/2}V_x(x)
\right]
\,,
\end{equation}
where
\begin{equation}
J(x)\equiv(E+Bx)^2-4C(Ax^2+Dx+F)
\,,
\end{equation}
and
\begin{equation}\label{K}
K(x)\equiv\Gamma x + BE - 2CD
\,.
\end{equation}
Note that in (\ref{R}) we have
\begin{equation}
V_x(x)\equiv \frac{\partial V}{\partial x}(x,y(x))\,,\mbox{~and~}
V_y(x)\equiv \frac{\partial V}{\partial y}(x,y(x))\,,
\end{equation}
with $y(x)$ obtained from (\ref{gcon}).
In the general case, at first it is not obvious how to directly deal with the second-order ordinary differential equation (\ref{ODE}) due to its intrinsic highly nonlinear nature.  Nonetheless, an integration factor for (\ref{ODE}) can be written as
\begin{equation}\label{IF}
I(x,\dot{x})= J^{-1/2}\dot{x}
\,,
\end{equation}
considerably symplifying the problem.  Indeed, after
multiplying both sides of (\ref{ODE}) by (\ref{IF}), we obtain a total derivative which can be readily integrated to
\begin{equation}\label{Energy}
\frac{\dot{x}^2}{2}
\left[
1+\frac{1}{4C^2} \left( \pm J^{-1/2}K - B \right)^2
\right] + \frac{V(x)}{m} = \frac{\cal E}{m}
\,,
\end{equation}
where $\cal E$ denotes a first real integration constant which can be physically interpreted as the total energy of the particle and
\begin{equation}
V(x)\equiv V(x,y(x))\,. 
\end{equation}
Finally, relation (\ref{Energy}) can be solved for $\dot{x}$, allowing a second integration and producing the implicit solution
\begin{equation}\label{xsol}
t =  \bigintssss_{x_0}^{x(t)}\,\left[\frac{m
\Big(
1+(\pm J^{-1/2}(u)K(u)-B)^2/4C^2
\Big)
}{2({\cal E}-V(u))}  \right]^{1/2}du
\,,
\end{equation}
with $x_0\equiv x(0)$ denoting another integration constant.   In fact, the inverse of relation (\ref{xsol}) gives $x$ as function of $t$ in terms of the arbitrary constants $x_0$ and $\cal E$ which can be used to adjust the initial conditions.  Similarly, by the same token we may obtain $y(t)$ from 
\begin{equation}\label{ysol}
t =  \bigintssss_{y_0}^{y(t)}\,\left[\frac{m
\Big(
1+(\pm \bar{J}^{-1/2}(u)\bar{K}(u)-B)^2/4A^2
\Big)
}{2({\cal E}-\bar{V}(u))}  \right]^{1/2}du
\,,
\end{equation}
with
\begin{equation}
\bar{J}(y)\equiv(D+By)^2-4A(Cy^2+Ey+F)
\,,
\end{equation}
\begin{equation}
\bar{K}(y)\equiv \Gamma y+BD-2AE
\end{equation}
and
\begin{equation}
\bar{V}(y)\equiv V(x(y),y)\,. 
\end{equation}
Once we have $x(t)$ and $y(t)$ it is immediate to obtain $z(t)$.  Indeed, we may use equation (\ref{z}) directly or, alternatively, the symmetrized version
\begin{equation}
z=-\frac{(m\ddot{x}+V_x)(Bx+2Cy+E)+(m\ddot{y}+V_y)(2Ax+By+D)}{2(2Ax+By+D)(Bx+2Cy+E)}
\,.
\end{equation}

In order to illustrate the previous ideas, we may consider the particular case of an elliptical path with semi-axes $a$ and $b$ given by
\begin{equation}\label{ellipseACF}
A=\,1/a^2,\,\,\,\,\,
C=1/b^2\,,\,\,\,\,\,
F=-1
\end{equation}
and
\begin{equation}\label{ellipseBDE}
B=D=E=0\,,
\end{equation}
which means equation (\ref{gcon}) reads
\begin{equation}
\frac{x^2}{a^2}+\frac{y^2}{b^2}=1\,.
\end{equation}
In terms of (\ref{ellipseACF}) and (\ref{ellipseBDE}), equations (\ref{P}) to (\ref{R}) are given by
\begin{equation}
P_\pm= \pm \frac{2}{b} \left( 1- \frac{x^2}{a^2} \right)^{1/2}
\left[ 1 + \frac{b^2x^2}{a^4(1-x^2/a^2)} \right]
\,,~~~\
Q_\pm=  \frac{\pm2bx}{a^4\left(1-x^2/a^2\right)^{3/2}}\,,
\end{equation}
\begin{equation}
R_\pm = \frac{2}{m} \left[  \pm \frac{1}{b}\left( 1 - \frac{x^2}{a^2} \right)^{1/2}\!\! V_x\, -\,\frac{xV_y}{a^2}\, \right]
\,,
\end{equation}
and then, after a convenient multiplication by $\pm ba^6(1-x^2/a^2)^{3/2}$, relation (\ref{ODE}) can be written as
the non-linear ordinary differential equation
\begin{equation}\label{ode}
\left[ (a^2-b^2)x^4 + a^2(b^2-2a^2)x^2+a^6 \right] \ddot{x} +a^2b^2 x\dot{x}^2=F(x)
\end{equation}
with
\begin{equation}
F(x) = am^{-1}(a^2-x^2)^{3/2}\left[ \mp a(a^2-x^2)^{1/2}V_x(x) + bxV_y(x)\right]
\,.
\end{equation}
By means of the integration factor (\ref{IF}), we may rewrite (\ref{ode}) as
\begin{equation}
\frac{d}{dt}\,\Big[
\frac{a^4+(b^2-a^2)x^2}{a^2-x^2}\dot{x}^2
+2a^2m^{-1}V
\Big]=0\,,
\end{equation}
and then perform the first integration obtaining the constant of motion
\begin{equation}
{\cal E}\equiv
\frac{m}{2a^2}
\Big[
\frac{a^4+(b^2-a^2)x^2}{a^2-x^2}
\Big]
\dot{x}^2\,+V
\end{equation}
which can be physically interpreted as the energy of the particle.  Corresponding to equations (\ref{xsol}) and (\ref{ysol}), for the present case we have 
\begin{equation}\label{xsol-ell}
t = 
\frac{\sqrt{2m}}{2a}
 \bigintssss_{x_0}^{x(t)}\,\left[
 \frac{
a^4+(b^2-a^2)u^2
}{({\cal E}-V(u))(a^2-u^2)}  
\right]^{1/2}du
\end{equation}
and
\begin{equation}\label{ysol-ell}
t = 
\frac{\sqrt{2m}}{2b}
 \bigintssss_{y_0}^{y(t)}\,\left[
 \frac{
b^4+(a^2-b^2)u^2
}{({\cal E}-\bar{V}(u))(b^2-u^2)}  
\right]^{1/2}du
\,.
\end{equation}
Depending on the specific form of the potential function, the last two equations can lead to well-known elliptic integrals \cite{GR}.

\section{Canonical Quantization}
After obtaining the general solution for the differential equations of motion, we turn now to the canonical quantization of the second-class system (\ref{L}).
This can be achieved by promoting the phase space variables to operators acting on a corresponding Hilbert space of wave functions and implementing the constraints as strong relations, i.e., $\chi_r\equiv0$. 
Since we are working with a constrained system, the naive prescription for 
obtaining the quantum operator commutators from the corresponding classical Poisson brackets does not work.  Clearly the PB algebra is inconsistent with the fact that the constraints (\ref{constraints}) should vanish along the classically allowed solutions.  In particular, the Hamilton equations cannot be consistently written in terms of PBs.  In order to circumvent those problems, we use instead the Dirac brackets (DB) algebra, which is tailor-made for second-class systems.  Given two arbitrary phase space functions $F$ and $G$, we define its Dirac bracket as \cite{Dirac, Sundermeyer:1982gv, Gitman:1990qh, Henneaux:1992ig}
\begin{equation}\label{DBd}
\{F,G\}^*\equiv \{F,G\} - \sum_{r,s=1}^{4}\{F,\chi_r\} \Delta^{rs}\{\chi_s,G\}
\,,
\end{equation}
where $\Delta^{rs}$ stands for the inverse of the constraints algebra matrix, that is,
\begin{equation}
\sum_{t=1}^4
\Delta^{rt}\{\chi_t,\chi_s\}=\delta^r_s\,.
\end{equation}
For the present case, the constraint relations (\ref{CPB14}) to (\ref{CPB34}) lead to
\begin{equation}\label{Delta}
\Delta^{rs} = mM^{-1}
\left(
\begin{array}{cccc}
0&-mM^{-1}\{\chi_3,\chi_4\}&mM^{-1}\{\chi_2,\chi_4\}&-1\\
mM^{-1}\{\chi_3,\chi_4\}&0&-1&0\\
-mM^{-1}\{\chi_2,\chi_4\}&1&0&0\\
1&0&0&0
\end{array}
\right)
\,,
\end{equation}
with $M=M(x,y)$, $\{\chi_2,\chi_4\}$ and $\{\chi_3,\chi_4\}$ respectively given by equations
(\ref{M}), (\ref{CPB24}) and (\ref{CPB34}).  By using (\ref{Delta}), the non-null DBs among the physical variables $(x,y,p_x,p_y)$ can be obtained from (\ref{DBd}) as
\begin{equation}\label{fDBsA}
\begin{split}
\{x,p_x\}^*&=M^{-1}\big(Bx+2Cy+E\big)^2
\,,\\
\{x,p_y\}^*&=\{y,p_x\}^*=-M^{-1}\big(2Ax+By+D\big)\big(Bx+2Cy+E\big)
\,,\\
\{y,p_y\}^*&=M^{-1}\big(2Ax+By+D\big)^2
\,,
\end{split}
\end{equation}
and
\begin{equation}\label{fDBsB}
\{p_x,p_y\}^*=M^{-1}\big[
(2Ap_x+Bp_y)(Bx+2Cy+E) - (Bp_x+2Cp_y)(2Ax+By+D)
\big]
\,.
\end{equation}
For completeness,
the non-null Dirac brackets involving the auxiliary variable $z$ are calculated in Appendix {\bf B}. 

A connection with the previously discussed main differential equation (\ref{ODE}) can be made here, by noticing that $\{p_x,p_y\}^*$ in equation (\ref{fDBsB}) above can be alternatively rewritten as
\begin{equation}\label{fDBsC}
\{p_x,p_y\}^*=M^{-1}\big[K(x)p_y-\bar{K}(y)p_x\big]
\,,
\end{equation}
with $K(x)$ given by equation (\ref{K}) and $\bar{K}(y)$ similarly defined as
\begin{equation}
\bar{K}(y)\equiv \Gamma y+BD-2AE\,.
\end{equation}
It is noteworthy that the general potential $V(x,y)$ present in (\ref{L}) does not show up in the fundamental Dirac brackets (\ref{fDBsA}) and (\ref{fDBsB}) above.

Now we pass to the corresponding quantum theory by promoting the phase space variables to operators satisfying commutation relations proportional to the Dirac brackets in equations (\ref{fDBsA}) and (\ref{fDBsB}).  More precisely, if $v$ and $w$ denote any given pair of phase space variables, we consider the two associated operators ${\hat v}={\cal F}(v)$ and ${\hat w} = {\cal F}(w)$ satisfying
\begin{equation}\label{QA}
[ \hat{v} , \hat{w} ]_- = i\hbar {\cal F}(\,\{ v,w \}^*)
\,,
\end{equation}
where $\cal F$ denotes the application which sends the phase space variables to the associated quantum operators.  In order to find a concrete realization $\cal F$ of the algebra (\ref{QA}) without operator ordering ambiguities, since $\hat x$ and $\hat y$ commute, we consider a Hilbert space of complex functions $\psi(x,y)$ of two real variables $x$ and $y$ and define the operators $\hat x$ and $\hat y$ as usual multiplication by $x$ and $y$.   
Thus the action of $\cal F$ on the RHS of equations (\ref{fDBsA}) is well defined.
Concerning the momentum operators $\hat{p}_x$ and $\hat{p}_y$, inspired on the unconstrained case, it is possible to find a representation of the Dirac algebra in terms of partial derivatives.  Indeed, if we define the action of $\cal F$ on (\ref{fDBsC}) as
\begin{equation}\label{pxpy}
[ \hat{p}_x,\hat{p}_y ]_-=i\hbar M^{-1}\big[K(x)\hat{p}_y-\bar{K}(y)\hat{p}_x\big]
\,,
\end{equation}
then the differential operator representation
\begin{equation}\label{px}
\hat{p}_x = i\hbar M^{-1}(Bx+2Cy+E)\big((2Ax+By+D)\partial_y-(Bx+2Cy+E)\partial_x\big)
\end{equation}
and
\begin{equation}\label{py}
\hat{p}_y = i\hbar M^{-1}(2Ax+By+D)\big((Bx+2Cy+E)\partial_x-(2Ax+By+D)\partial_y\big)
\end{equation}
does the job.  To prove this assertion, first notice that from the differential representation in equations (\ref{px}) and (\ref{py}) above we may write
\begin{equation}\label{hatpxhatpy}
\hat{p}_x\hat{p}_y=i\hbar M^{-1}\Big[ \Xi_{xy}^{ACDE} +  \Theta_{xy}^{ACDE} \Big]
\end{equation}
with
\begin{equation}
\Xi_{xy}^{ACDE} = \big(2A(Bx+2Cy+E)-B(2Ax+By+D)\big)\hat{p_x}
\end{equation}
and
\begin{eqnarray}
\Theta_{xy}^{ACDE}=(2Ax+By+D)\hat{p}_x\Big((Bx+2Cy+E)\partial_x-(2Ax+By+D)\partial_y\Big)
\phantom{(2Ax)}\nonumber\\
-M^{-1}(Bx+2Cy+E)\Big((2Ax+By+D)M_y -(Bx+2Cy+E)M_x\Big)\hat{p}_y
\,.
\end{eqnarray}
Next, observe that due to the symmetry properties
\begin{equation}
T_x\hat{p}_x=-T_y\hat{p}_y
\end{equation}
and
\begin{equation}
\Theta_{xy}^{ACDE}=\Theta_{yx}^{CAED}
\,,
\end{equation}
the quantity $\Theta_{xy}^{ACDE}$ does not contribute to the commutator and from (\ref{hatpxhatpy}) we have 
\begin{eqnarray}
[ \hat{p}_x,\hat{p}_y ]_-&=&i\hbar M^{-1}\Big[ \Xi_{xy}^{ACDE} -  \Xi_{yx}^{CAED} \Big]\nonumber\\
&=&i\hbar M^{-1}\big[K(x)\hat{p}_y-\bar{K}(y)\hat{p}_x\big]
\,,
\end{eqnarray}
coinciding exactly with (\ref{pxpy}), which is what we wanted to prove.  Furthermore, using (\ref{px}) and (\ref{py}), it is straightforward to show that
\begin{equation}\label{QfDBsA}
\begin{split}
[x,\hat{p}_x]_-&=i\hbar M^{-1}\big(Bx+2Cy+E\big)^2
\,,\\
[x,\hat{p}_y]_-&=[y,\hat{p}_x]_-=-i\hbar M^{-1}\big(2Ax+By+D\big)\big(Bx+2Cy+E\big)
\,,\\
[y,\hat{p}_y]_-&=i\hbar M^{-1}\big(2Ax+By+D\big)^2
\,,
\end{split}
\end{equation}
which correspond to the quantum commutators generated by the action $\cal F$ on the Dirac brackets (\ref{fDBsA}).
For instance,
\begin{eqnarray}
[x,\hat{p}_x]_-&=&x\hat{p}_x-\hat{p}_x x \nonumber\\
&=&i\hbar\Big[xM^{-1}(Bx+2Cy+E)\Big((2Ax+By+D)\partial_y-(Bx+2Cy+E)\partial_x\Big)  \nonumber\\&&
-M^{-1}T_y\Big((2Ax+By+D)\partial_y-(Bx+2Cy+E)\partial_x\Big)x \Big]  \nonumber\\
&=&-i\hbar M^{-1}(Bx+2Cy+E) (-Bx-2Cy-E)  \nonumber\\
&=&i\hbar M^{-1}(Bx+2Cy+E)^2
\,,
\end{eqnarray}
and so on and so forth similarly for the other relations in (\ref{QfDBsA}).  Hence, we have explicitly obtained a differential operator representation at quantum level for the Dirac brackets.  
To conclude the quantization process, we enforce all constraint relations as strong identities and write down the quantum Hamiltonian operator as
\begin{equation}
\hat{H}=\frac{1}{2m}(\hat{p}_x^2+\hat{p}_y^2)+V(x,y)
\end{equation}
acting on the Hilbert space of complex functions $\psi(x,y)$, with $\hat{p}_x$ and $\hat{p}_y$ given by equations (\ref{px}) and (\ref{py}).

\section{Conclusion and Final Remarks}
We have discussed the general conic path second-class system with an arbitrary open potential function in detail, providing its classical solution in full length and performing its canonical quantization via Dirac brackets.  For those accounts, the system proposed and discussed here considerably enlarges and complements the two previously published works \cite{Barbosa:2018dmb} and \cite{Barbosa:2018bng}.  In order to handle the corresponding non-linear second order differential equations, an explicit integration factor was exhibited allowing its order reduction and subsequent integration. We have also obtained the complete PB relations among all constraints, as well as the Lagrange multiplier function, in a neat closed form clarifying the dynamical evolution of the constrained system in phase space.  Considering our purposes, in this letter, we have focused on the second-class system itself.  In spite of that, it is noteworth to stress that the system studied here can also be described in a gauge invariant way along the lines of references \cite{Barbosa:2018dmb, Barbosa:2018bng, deOliveira:2019eva}.  Indeed, it has been shown in the cited references that the BRST symmetry can be straightforward implemented allowing for the introduction of anticommuting ghost variables and for a BFV\footnote{Batalin–Fradkin–Vilkovisky \cite{Fradkin:1975cq, Batalin:1977pb}.} functional quantization approach. One can then benefit from the general solution obtained here in order to better understand those BRST/BFV aspects and to strength the comparative analysis with specific field theory models as in references \cite{Nemeschansky:1987xb, Gupta:2009dy, Shukla:2014spa}.  The conic path system has recently inspired the construction of a gauge-invariant prototypical mechanical system \cite{Pandey:2021myh, Thibes:2020yux} with direct application to quantum field theory models such as the Lorentz symmetry breaking bumblebee model \cite{Pandey:2021myh} and the nonlinear sigma model \cite{Thibes:2020yux}.  The existence of the general solution presented here should also bring important clues for a deeper analysis of that recently proposed prototypical system.

\appendix
\section*{Appendix A - Constraint Poisson Brackets}
We present here the technical details for obtaining equations (\ref{chi3lambda}) and (\ref{chi4lambda}) as well as the  explicit form for the corresponding functions $\Omega$ and $F$.

To obtain $\{\chi_3,\lambda\}$ from equations (\ref{constraints}) and (\ref{lambda}), notice that if we define
\begin{equation}
T_x\equiv (2Ax+By+D)\,,~~~T_y\equiv (Bx+2Cy+E)\,,
\end{equation}
\begin{equation}
X\equiv T_x  (V_{xx}+8Az)+ T_y(V_{xy}+4Bz)+3BV_y+6AV_x
\end{equation}
and
\begin{equation}
Y\equiv T_x(V_{xy}+4Bz)+T_y(V_{yy}+8Cz) +6CV_y + 3BV_x
\,,
\end{equation}
we may write
\begin{equation}
\{\chi_3,\lambda\}=
-\frac{1}{m^2M^2}\left[
p_x \Omega^{x} + p_y \Omega^{y}
\right]
\,,
\end{equation}
with
\begin{equation}
\Omega^x = M(-T_x X_x + 2AX + BY) - T_y X_y +X(T_xM_x+T_y M_y)\,,
\end{equation}
and
\begin{equation}
\Omega^y = M(-T_xY_x+BX-T_yY_y+2CY) + Y(T_xM_x+T_yM_y)\,.
\end{equation}
Then, an explicit calculation gives
\begin{eqnarray}
\Omega^x &=& \Big[ 6AT_x^2 + 4B T_xT_y + (2A+4C)T_y^2 \Big] X + BMY\nonumber\\
&&-M \Big[ T_x(8AV_{xx}+4BV_{xy}+(16A^2+4B^2)z + T_x V_{xxx} + T_y V_{xxy})
\,,
\nonumber\\
&&+T_y ((2C+6A)V_{xy} +  8B(A+C)z + T_x V_{xxy} + T_y V_{xyy} + BV_{xx} + 3B V_{yy}) \Big]
\end{eqnarray}\label{66}
and
\begin{eqnarray}\label{67}
\Omega^y &=& \Big[ (4A+2C)T_x^2 + 4B T_xT_y + 6CT_y^2 \Big] Y + BMX\nonumber\\
&&- M \Big[ T_x ((2A+6C)V_{xy} +  8B(A+C)z + T_x V_{xxy} + T_y V_{xyy} + BV_{yy} + 3B V_{xx})
\nonumber\\
&&+ T_y(8CV_{yy}+4BV_{xy}+(16C^2+4B^2)z + T_y V_{yyy} + T_x V_{xyy})  \Big]
\,,
\end{eqnarray}
which corresponds to the expressions for $\Omega^{x}(x,y,z)$ and $\Omega^{y}(x,y,z)$ within equation (\ref{chi3lambda}).

Concerning $\{\chi_4,\lambda\}$, note that if we define further
\begin{equation}
P\equiv Ap_x^2+Bp_xp_y+Cp_y^2\,,
\end{equation}
we may write
\begin{equation}
\{\chi_4,\lambda\}=\frac{2}{m^2}\{P,\lambda\}-\frac{z}{m}\{ M,\lambda \} -\frac{1}{m} \{N,\lambda\}
\end{equation}
with
\begin{equation}
\frac{2}{m^2}\{P,\lambda\}=p_x^2F_{(2)}^{xx}+p_xp_yF_{(2)}^{xy}+p_y^2F_{(2)}^{yy}
\,,
\end{equation}
where
\begin{equation}
F_{(2)}^{xx}=
\frac{2}{m^3M^2}\Big[
(2AX_x+BX_y)M-2X\Big( (4A^2+B^2)T_x+2B(A+C)T_y \Big) \Big]
\,,
\end{equation}
\begin{equation}
F_{(2)}^{yy}=
\frac{2}{m^3M^2}\Big[
(BY_x+2CY_y)M-2Y\Big( 2B(A+C)T_x + (4C^2+B^2)T_y \Big) \Big]
\end{equation}
and
\begin{eqnarray}
F_{(2)}^{xy}&=&
\frac{2}{m^3M^2}\Big[
(2AY_x+BX_x+BY_y+2CX_y)M-2(BX+2AY)(2AT_x+BT_y)
\nonumber\\&&
-2(BY+2CX)(BT_x+2CT_y)\Big]
\end{eqnarray}
correspond to the functions defined in equation (\ref{chi4lambda}).  Furthermore, note that the remaining momenta-independent part of $\{\chi_4,\lambda\}$ is given by 
\begin{eqnarray}
F_{(0)}&=&-\frac{z}{m}\{ M,\lambda \} -\frac{1}{m} \{N,\lambda\} \nonumber\\
&=&\frac{1}{m^2M}
\Big[
X(zM_x+N_x)+Y(zM_y+N_y)
\Big] 
\end{eqnarray}
corresponding to $F_{(0)}$ as defined in (\ref{chi4lambda}).

\section*{Appendix B - DBs for the auxiliary variable $z$}
Concerning the auxiliary variable $z$ introduced in (\ref{L}) to implement the restriction (\ref{gcon}), we have the non-null DBs
\begin{equation}
\begin{split}
\{x,z\}^*&=2m^{-1}M^{-1}\big(Bx+2Cy+E\big)
\{p_x,p_y\}^*\,,\\
\{y,z\}^*&=2m^{-1}M^{-1}(2Ax+By+D)
\{p_y,p_x\}^*\,,
\end{split}
\end{equation}
and
\begin{equation}\label{zpxpy}
\begin{split}
\{z,p_x\}^*&=-2m^{-1}
\left(2Cp_y+Bp_x\right)
\{p_x,p_y\}^*
+M^{-2}(Bx+2Cy+E)
\Big[ V_x \bar{K}(y)-V_yK(x)
\\&
~~~
+\big((2Ax+By+D)^2
-(Bx+2Cy+E)^2\big)\big(V_{xy}+2zB\big)
\\&
~~~
-(2Ax+By+D)(Bx+2Cy+E)\big(V_{xx}-V_{yy}+2A-2C\big) \Big]
\,,\\
\{z,p_y\}^*&=2m^{-1}
\left(2Ap_x+Bp_y\right)
\{p_x,p_y\}^*
-M^{-2}(2Ax+By+D)
\Big[ V_x \bar{K}(y)-V_yK(x)
\\&
~~~
+\big((2Ax+By+D)^2
-(Bx+2Cy+E)^2\big)\big(V_{xy}+2zB\big)
\\&
~~~
-(2Ax+By+D)(Bx+2Cy+E)\big(V_{xx}-V_{yy}+2A-2C\big) \Big]
\,,
\end{split}
\end{equation}
which also follow from the definition (\ref{DBd}).
As we can see, the derivatives of the potential function $V(x,y)$ show up only in the Dirac brackets among $z$ and the momenta in (\ref{zpxpy}).  Of course the DB between any phase space function and $p_z$ is null, as the latter constitutes one of the constraints (\ref{constraints}) in itself.

\end{document}